**Tuning the Magnetization of Manganese (II) Carbonate by Intracrystalline Amino Acids**

Arad Lang[1], Iryna Polishchuk[1], Giorgia Confalonieri[2], Catherine Dejoie[2], Ariel Maniv[3], Daniel Potashnikov[4], El'ad N. Caspi[3], Boaz Pokroy[1,5]∗.

[1] Department of Materials Science and Engineering and The Russell Berrie Nanotechnology Institute, Technion – Israel Institute of Technology, 32000 Haifa, Israel.

[2] ESRF – The European Synchrotron Radiation Facility, CS 40220, 38043 Grenoble Cedex 9, France.

[3] Physics Department, Nuclear Research Centre, Negev, P.O. Box 9001, Beer-Sheva 84190, Israel.

[4] Israel Atomic Energy Commission, P.O. Box 7061, Tel-Aviv 61070, Israel

[5] The Nancy and Stephen Grand Technion Energy Program, Technion − Israel Institute of Technology, Haifa 3200003, Israel

E-mail: bpokroy@technion.ac.il.

**Abstract**

Incorporation of organic molecules into the lattice of inorganic crystalline hosts is a common phenomenon in biomineralization, and has been shown to alter various properties of the crystalline host. Taking this phenomenon as our source of inspiration, we show herein that incorporation of specific single amino acids into the lattice of manganese (II) carbonate strongly alters its inherent magnetic properties. At room temperature, the magnetic susceptibility of the amino-acid-incorporating paramagnetic $MnCO_3$ decreases, following a simple rule of mixtures. When cooled below the Néel temperature, however, the opposite trend is observed, namely an increase in magnetic susceptibility measured in a high magnetic field. Such an increase, accompanied by a drastic change in the Néel phase transformation temperature, results from a decrease in $MnCO_3$ orbital overlapping and the weakening of superexchange interactions. To our knowledge, this is the first time that the magnetic properties of a host crystal have been tuned via the incorporation of amino acids.

# 1. Introduction

Inorganic crystals grown by organisms via biomineralization processes demonstrate fascinating strategies of crystal growth.[1] Their structures, from the atomic to the micron scale, are tailored to serve specific functions such as mechanical (mastication,[2,3] shielding,[4,5] support,[6,7]) optical[8] and magnetic.[9–13] One widespread strategy of biogenic crystal growth is the incorporation of organic molecules into the lattice of an inorganic crystalline host,[14–16] a procedure that has been shown also to induce lattice distortions in the host biocrystals.[17] This latter biostrategy has been translated, for example, to synthetic calcium carbonate ($CaCO_3$) by the incorporation of block copolymers,[18] polymeric micelles,[19,20] hydrogels,[21–23] dyes,[24] drugs,[25] peptides,[15,26,27] and monosaccharaides,[28] as well as of single amino acids (AAs).[29–31]

Various single AAs have also been incorporated into a variety of other synthetic semiconductors, including zinc oxide (ZnO),[32,33] copper oxide ($Cu_2O$)[34] and methylammonium lead bromide ($MAPbBr_3$)[35]. Such incorporation not only induces lattice distortions, but also increases the optical band gap of the crystalline host. This blue shift originates from the insulating nature of the AAs, which act as potential barriers inside the semiconducting matrix, and induces a quantum confinement effect.[36]

In this study, we aimed to determine whether incorporation of an AA into a magnetic crystalline host influences its magnetic properties. To this end we chose to study the magnetic properties of manganese (II) carbonate (MCO, rhodochrosite), which possesses the same crystal structure as that of calcite (*R-3c* space group)[37]. MCO is paramagnetic (PM) at room temperature and undergoes transformation to its antiferromagnetic (AFM) phase at a Néel temperature ($T_N$) of about 32 K.[38] It belongs to a family of materials, which while in their AFM phase, exhibit weak ferromagnetism (WFM). This is owing to spin canting, in which the spin vectors of the Mn atoms are tilted out of the *ab* plane of the crystal, resulting in a weak ferromagnetic response (also known as the Dzyaloshinskii–Moriya interaction, DMI).[39–42] MCO can be utilized in a variety of applications, including supercapacitors,[43–45] lithium-ion battery anodes,[46–48] plant fertilizers,[49] light emitters,[50] hematinic agents,[51] and contrast enhancing agents for magnetic resonance imaging (MRI).[52,53]

## 2. Results

MCO was synthesized by the use of both room-temperature drop-by-drop (DD) and hydrothermal (HT) methods, in the presence of the identical molar concentration (80 mM) in solutions of four common L-AAs: aspartic acid (Asp), glutamic acid (Glu), glycine (Gly) and cysteine (Cys). These AAs were selected because they are the ones with the most significant levels of incorporation into calcite (for more details, see Experimental section).[29,31,54] Cys was used only in the DD method, since it tends to decompose at high temperatures. **Figure 1** presents the change in magnetic susceptibility of AA-incorporated MCO compared to that of pure MCO (see magnetic moment *vs.* magnetic field plots (M−H) presented in **Figure S1**). The magnetic susceptibility ($\chi$) was calculated as the slope of the linear part of the normalized measured magnetization (*m*) in relation to the external magnetic field (*H*), i.e.:

$$\chi \equiv \frac{\partial m}{\partial H} \quad \left[\frac{emu}{Oe \cdot gr}\right]. \tag{1}$$

**Figure 1** shows, interestingly, that the magnetic susceptibility of MCO was indeed altered owing to AA incorporation. This alteration was more pronounced in the HT-grown samples than in the DD-grown ones, even though the AA concentration in solution was identical in both cases. This clear observation provided an indication of intracrystalline incorporation, since the incorporation level after HT synthesis is higher also in the case of calcite.[31] Even more striking were the opposite effects of AAs incorporation at the two different temperatures: while the magnetic susceptibility to incorporation decreased at room temperature, it increased dramatically at 2 K. An exception was observed in the case of Cys, whose magnetic susceptibility at both temperatures was decreased.

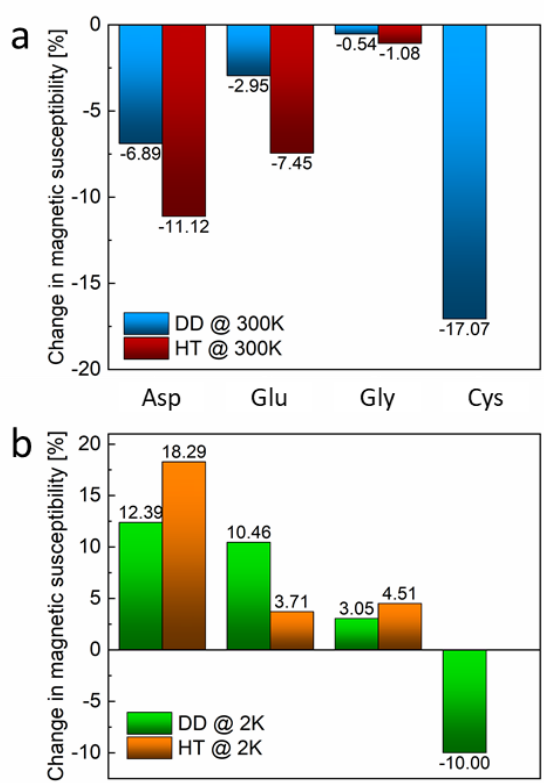

**Figure 1. Changes in magnetic susceptibility of AA-incorporated MCO.** Changes in the magnetic susceptibilities of MCO grown via HT and DD methods, in the presence of the same concentration (80 mM) of selected L-AAs, measured at (a) 300 K and (b)

2 K. Changes were calculated relative to pure MCO. The maximum relative error is 0.05% of the presented values.

Since both Asp and Cys showed the strongest effects on the magnetic susceptibility of MCO (**Figure 1**), we proceeded to further study the effects of various concentrations of these two AAs. To this end we added Asp or Cys into the solution at concentrations of 20, 50 and 80 mM. The exact amount of the incorporated AA was determined by amino acid analysis (AAA). The amount of incorporated AA was found to increase with its increasing concentration in the growth solution, reaching the highest level of ~ 8 mol % in the case of Asp (**Figure S2**). Interestingly, the incorporation level of Asp in the MCO lattice was of the same order of magnitude as that observed in the case of calcite,[31] and as expected, the incorporation levels due to higher temperature and pressure achieved via the HT growth method were higher than those achieved via DD.

We then used synchrotron high-resolution powder X-ray diffraction (HR-PXRD) to detect changes in MCO structure and microstructure caused by the incorporation of Asp. **Figure 2a** depicts representative interplanar spacings of the (104) plane, which is the most intense reflection of MCO (see **Figure S3**), for both the pure and the Asp-incorporated MCO. The $d$-spacings of the Asp-incorporated sample were larger than those of the control sample. This relative expansion is a known fingerprint for the incorporation of organic molecules.[29,32,55] Moreover, following mild thermal annealing (200°C for 4 h, in vacuum), which induces decomposition of the incorporated organics, the diffraction peaks shifted to smaller $d$-values. This is also a fingerprint, this time for incorporation of the organics within the crystalline lattice.[29,32] Furthermore, **Figure 2a** clearly demonstrates a significant broadening of the diffraction peaks upon Asp incorporation, suggesting a decrease in the crystallite size of the material. This suggestion is further supported by the high-resolution scanning electron microscopy (HR-SEM) images of the crystals (see **Figure S4**). Additional evidence for lattice incorporation came from behavior of the coherence length upon annealing, calculated via the Lorentzian contribution to the FWHM of the diffraction peaks (**Figure S5**).[56] Whereas after mild heat treatment the coherence length of the control samples was slightly increased (because of crystal growth due to annealing), it decreased in the case of Asp-incorporated crystals. This decrease was due to the creation of new defects upon organic degradation.[56] In the case of Cys, HR-PXRD revealed the formation of an additional phase (α-$Mn_2O_3$, bixbyite), whose amount increased with the increase of incorporated Cys in the crystal (see **Figure S11**). The decrease in magnetic susceptibility of Cys-incorporated MCO, observed both at 300 K and at 2 K (**Figure 1** and **Figure S11d**), stems from the formation of this second phase (for more details, see SI).

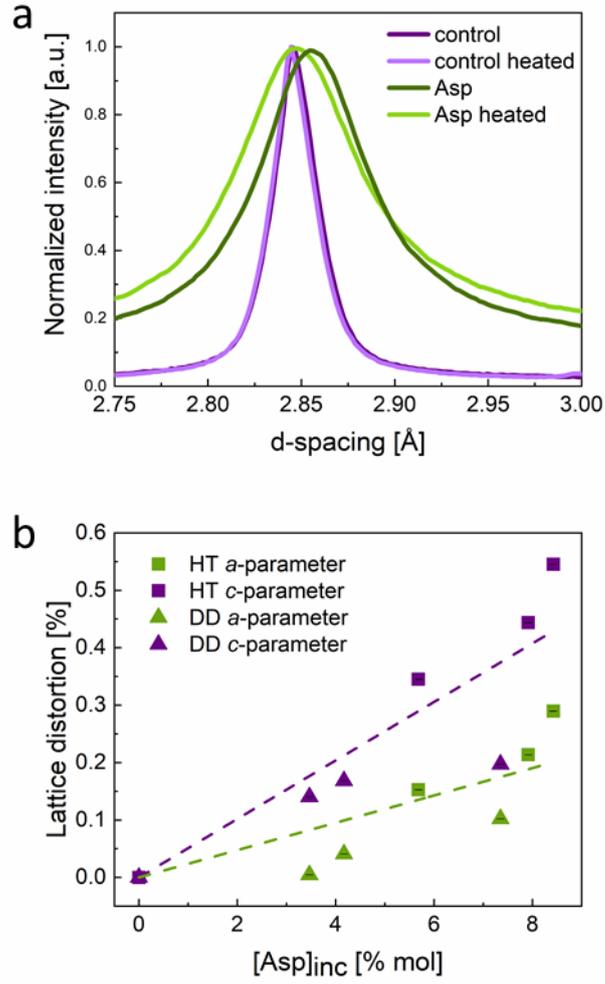

**Figure 2. Lattice distortions in Asp-incorporated MCO.** a) The (104) reflection of both pure and Asp-incorporated MCO, before and after thermal annealing. b) Lattice distortions in *a*- and *c*-parameters of MCO as a function of the amount of incorporated Asp. Error bars are shorter than the data points.

We used Rietveld refinement to accurately calculate the lattice parameters (both *a* and *c*) of each sample, and calculated the lattice distortion according to the equation:

$$\frac{\Delta a}{a} = \frac{a_A - a_C}{a_C} \qquad \frac{\Delta c}{c} = \frac{c_A - c_C}{c_C} \qquad (2)$$

where $a_A$, $c_A$ are the lattice parameters of the Asp-incorporated samples and $a_C$, $c_C$ are those of the control sample. As in calcite,[29] the MCO distortion was positive, anisotropic, and its magnitude increased with the amount of incorporated Asp (as measured here by AAA). In particular, on comparison of the lattice distortions induced by similar amounts of Asp in MCO and in calcite grown via the HT method, the measured distortion of the *c* lattice parameter was found to be similar to that of calcite.[31] In addition, and as also with calcite,[31] the HT method allowed the incorporation of higher AA levels into the lattice, and the higher the incorporated concentration the higher the lattice distortions (**Figure 2**).

To further study the effects of AAs incorporation on the crystal structure and on the magnetic structure of MCO, we employed neutron powder diffraction (NPD). The structural parameters obtained on NPD were in good agreement with those obtained by HR-PXRD (**Figure S6a**). More specifically, NPD also revealed anisotropic lattice distortions (in which the distortions along the *c* lattice parameter were greater than those along the *a* lattice parameter), as well as broadening of the diffraction peaks due to Asp incorporation (**Figure S6c**). One distinct feature shown by NPD was the dramatic increase in background intensity for the Asp-incorporated samples compared to that of the control. This is probably due to the presence in Asp of hydrogen atoms, which are known to increase the background intensity in NPD patterns.[57]

Following the above structural analysis, magnetization of the MCO samples was measured utilizing an MPMS3 SQUID magnetometer (Quantum Design). **Figure 3a,c** depicts the magnetization of both pure and Asp-incorporated HT-grown MCO as a function of the applied magnetic field (between −1 and 1 T) The M−H plots measured at room temperature (**Figure 3a**) were characteristic of a PM phase while those measured at 2 K fit a canted AFM phase[58,59] (**Figure 3c**). Similar results were obtained for the DD-grown samples (for the complete curves, see **Figure S7**). Both the extent of decrease in magnetic susceptibility at room temperature and its increase at 2 K were found to depend on the concentration of incorporated Asp (**Figure 3b,d**). Note that the samples demonstrated the same trend at the higher magnetic field between −7 and 7 T as that between −1 and 1 T (see **Figure S8**).

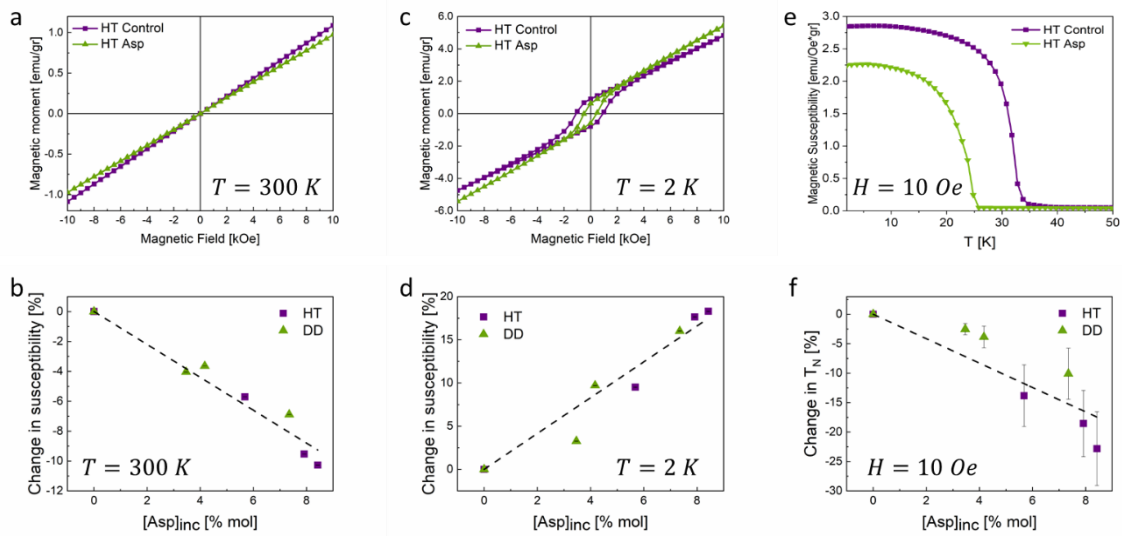

**Figure 3. Changes in magnetic properties of MCO due to incorporation of Asp**. (a, b) Decrease in magnetic susceptibility measured at room temperature. (c, d) Increase in magnetic susceptibility measured at 2 K. (e, f) Decrease in $T_N$ measured during heating at zero magnetic field relative to the control sample. In (b) and (d) the error bars are smaller than the data points.

Interestingly, along with the induced changes in magnetic susceptibility, Asp incorporation significantly affected the Néel temperature ($T_N$) of MCO, namely the temperature of magnetic phase transformation (**Figure 3e,f**). A decrease of over 20% in $T_N$ (by ~6 K) was achieved at the maximum concentration of incorporated Asp (~8 mol %). Similarly to the changes in $\chi$ upon Asp incorporation, the extent of change in $T_N$ was dependent on the amount of incorporated Asp rather than on the method of synthesis.

To further study the effects of Asp incorporation on the magnetic properties of MCO we repeated our NPD measurements below $T_N$, at a temperature of 3 K. Additional diffraction peaks were observed due to the magnetic phase transformation (**Figure S7b,d**). The quality of these NPD data did not allow for determination of the WFM known to exist in MCO[41,60] and also observed in this study (**Figure 3c**). However, the AFM portion of the magnetic ordering was easily determined using Rietveld refinement when we used a model exhibiting *C2′/c′* symmetry (#15.89)[61] (**Figure 4**). We first extracted the magnetic moments for both of the control samples derived via the synthesis routes DD and HT, and found that these magnetic moment values were in an excellent agreement with known published data (~4.2 $\mu_B$).[42] Similar extractions in the case of the Asp-incorporated samples, however, resulted in significant reductions in the ordered magnetic moments.

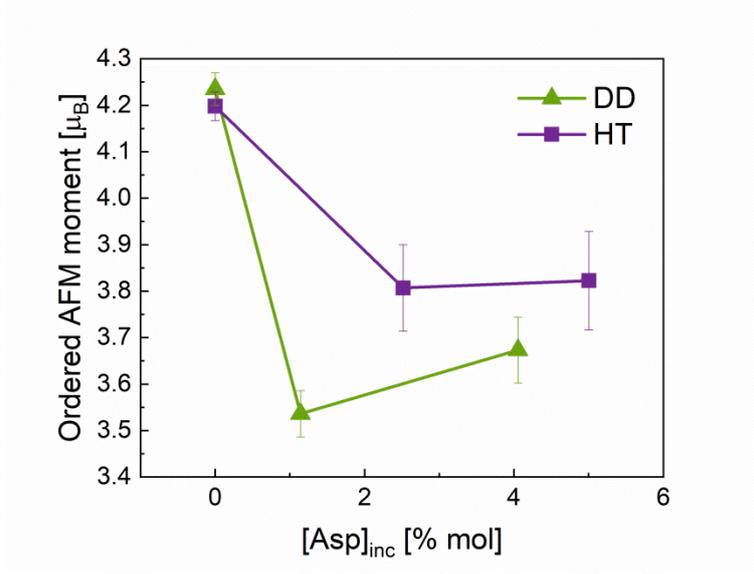

**Figure 4**. **Changes in antiferromagnetic moment of MCO.** Ordered AFM moment of MCO determined via Rietveld refinement utilizing the magnetic symmetry model *C2'/c'* (#15.89) in the basis (a,b,c)(0,0,0) of the parent space group *R-3c* to the NPD data.

## 3. Discussion

At room temperature, the magnetic susceptibility of MCO decreased with increasing Asp concentration (**Figure 3a−c**). This was probably due to the creation of a composite material. To verify this assumption, we measured the susceptibility of a pure Asp powder (see **Figure S10**), and used the rule of mixtures to calculate the expected susceptibility of the composite:

$$\chi_{total} = w_{Asp}\chi_{Asp} + (1-w_{Asp})\chi_{MCO}, \tag{3}$$

where $w_{Asp}$ is the weight fraction of incorporated Asp (as obtained by AAA) and $\chi_{Asp}$ and $\chi_{MCO}$ are the susceptibilities of pure Asp and pure MCO (control samples), respectively. **Figure 5a** indeed demonstrates a good correlation between the measured data and the calculated prediction.

In contrast to PM susceptibility at room temperature, the increase in MCO susceptibility measured at T=2 K cannot be explained by a simple rule of mixtures. To explain this effect, we first need to understand what types of magnetic interaction exists in MCO. At low temperatures (below the $T_N$), MCO is an asymmetric superexchange magnet (i.e., it demonstrates the so-called Dzyaloshinskii-Moriya interaction, DMI).[41,42] According to this model, the spin vectors of the magnetic cations interact via bridging of the non-magnetic anions[62]—an interaction known to be strongly dependent on their interatomic distance.[63] In the case of low-temperature MCO, the AFM ordering of the $Mn^{+2}$ cations occurs due to interaction through the oxygen atoms, while the alternating orientation of the $CO_3^{-2}$ anions induces an out-of-plane spin canting, which leads to the observed WFM.[41,42,58] Probably, the positive lattice distortion (**Figure 2b**) induced by the Asp incorporation weakens the DMI by increasing the Mn-O distance. Hence, the crystal becomes more "magnetically compliant", i.e., the low temperature susceptibility increases and the WFM hysteresis loop narrows more rapidly (**Figure 3c**). Moreover, this reduced interaction allows the magnetic ordering to break at lower temperatures, resulting in a decrease of the $T_N$ (**Figure 3e,f**). To verify this hypothesis, we used cryo-HR-PXRD to measure the lattice parameters of MCO at 5 K. **Figure 5b** presents the magnetic susceptibility (taken at the linear region of the M−H plot measured at 2 K) as a function of the Mn-O inter-atomic distance calculated using the previously reported equations developed for isostructural calcite[64] (see full HR-PXRD pattern in **Figure S11** and **Table S3**). The existing strong correlation ($R^2 = 0.96$) indeed suggested that the increase in bond length, which reduces the Mn-3d and O-2p orbitals, reduces the magnetic superexchange interaction in the crystals, thus resulting in a higher susceptibility and a narrower magnetic hysteresis loop.

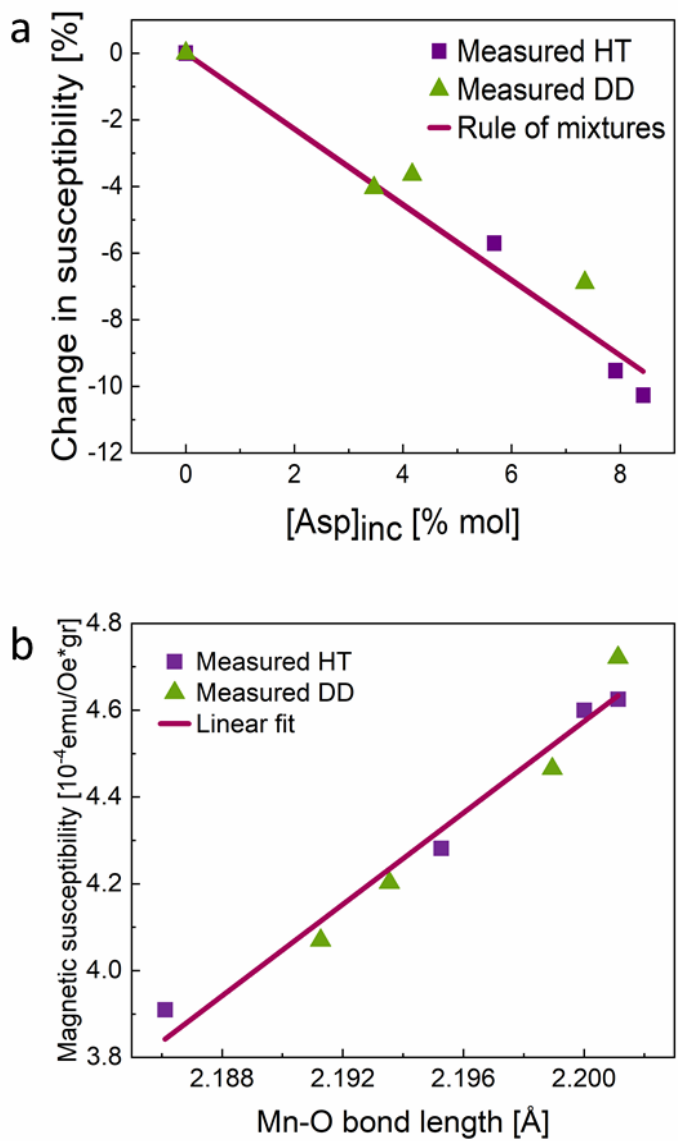

**Figure 5. Suggestions for the observed changes in magnetic susceptibilities.** (a) Decrease at room temperature: comparison between the measured susceptibilities and the calculated prediction using equation (3). (b) Increase at low temperature: measured susceptibilities as a function of the Mn-O distance in the crystal. Error bars are smaller than data points.

The significant reduction in the ordered magnetic moment of MCO as a function of Asp incorporation, as observed by NPD, cannot be easily explained in terms of weakening of the DMI. Such a reduction might be correlated to a possible change in lattice symmetry driven by this incorporation. If such a change in symmetry indeed exists, we could hypothesize that the chosen refinement model is not ideal, and that a reduction in symmetry should be considered and sought. An additional indication of such a symmetry reduction on the crystallographic level can be demonstrated by refining the oxygen positions of the *C2'/c'* space group (**Figure 6**). When these atomic positions were refined to fit the diffraction data of the DD sample or of the control, identical atomic positions emerged for both O1 and O2. Moreover, this position value (~0.28) was the value expected when the parent *R-3c* symmetry is used to refine the data. In practice, therefore, no reduction in crystallographic symmetry was expected for these samples. However, when the same refinement process was repeated for the HT samples collected at *both* RT and 3 K, a clear split was observed between the O1 and the O2 positions. This splitting became more prominent upon incorporation of higher amounts of Asp (**Figure 6**). Those results clearly suggested that incorporation of Asp under the HT process, which is known to lead to higher incorporation levels of AAs, causes a reduction in the crystallographic symmetry.

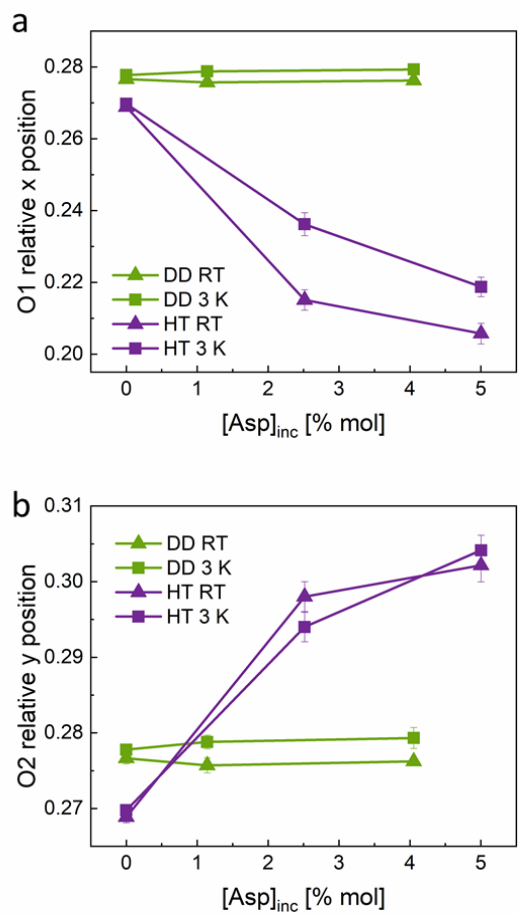

**Figure 6. Changes in the relative oxygen position due to Asp incorporation.** Relative oxygen position as a function of Asp concentration, as determined by application of Rietveld refinement to the NPD data by using the *C2'/c'* magnetic space group in the (a,b,c)(0,0,0) basis of the parent *R-3c* space group. In this model, O1 is positioned in (x~0.25, 0, ¼) (a) and O2 is positioned in (0, y~0.25, ¼) (b).

## 4. Conclusion

In this study we show for the first time that incorporation of AAs into the crystal lattice of manganese (II) carbonate induces significant changes in its magnetic susceptibility. This phenomenon depends on the magnetic ordering of MCO: at room temperature MCO is inherently paramagnetic and its susceptibility upon incorporation of AAs decreases by the simple rule of mixture. At low temperatures MCO is a canted antiferromagnet, which exhibits weak ferromagnetism. Upon AA incorporation it becomes more magnetically compliant owing to weakening of the superexchange interaction, induced by AAs-incorporation-governed lattice expansion. Moreover, as a result of the AAs incorporation, the transformation temperature of these two magnetic states is decreased.

This study demonstrates the feasibility of tuning the magnetic properties of a crystalline host via a bioinspired route, namely incorporation of AAs. It therefore opens a new avenue of research in the realm of magnetic materials and paves the way to a novel approach to the tailoring of their magnetic properties.

## 5. Methods

*Materials*: Stock solutions of 0.2 M $Mn^{+2}$ and $CO_3^{-2}$ were prepared by dissolving, respectively, 19.79 g of manganese chloride tetrahydrate ($MnCl_2*4H_2O$, Fisher Scientific) and 10.60 g of sodium carbonate ($Na_2CO_3$, Mercury) in 500 ml of DI $H_2O$. Different amounts of L-Aspartic acid (Asp, Sigma-Aldrich), L-Glutamic acid (Gly, Alfa Aesar), L-Glycine (Gly, Sigma-Aldrich) and L-Cysteine (Cys, Sigma-Aldrich) were used.

*$MnCO_3$ synthesis*: Synthesis was carried out via two routes: (i) Room temperature Drop-by-Drop (DD) method:[29] Following transfer of 50 ml of $CO_3^{-2}$ solution to a glass beaker, an appropriate amount of an L-AA was added to the solution (20, 50, 80 mM of Asp or Cys, 80 mM of Gly or Glu), and stirring was continued until the AA was completely dissolved. Meanwhile, 50 ml of $Mn^{+2}$ solution was loaded into a plastic syringe, which was placed in a syringe pump. MCO precipitation was initiated by the slow addition of drops of $Mn^{+2}$ solution into the $CO_3^{-2}$ solution at controlled dropping and stirring rates (4 ml min$^{-1}$, 400 rpm), and the solution was then stirred for 1 h after the addition was completed. The precipitated powder was washed, centrifuged several times in DI $H_2O$ and once with ethanol, and dried overnight at 50ºC under vacuum.

(ii) Hydrothermal (HT) method:[28,31] 50 ml of $CO_3^{-2}$ solution was transferred to a beaker. An appropriate amount of an L-AA was added to the solution (the same amounts and concentrations and the same stirring procedure as described for the DD route). This was followed by the rapid addition of 50 ml of $Mn^{+2}$ solution to the beaker, which was immediately transferred to a pre-heated autoclave. The autoclave was sealed and kept at 134ºC (~3 bar) for 2 h, after which it was allowed to cool to room temperature. The resulting powder was washed and dried as for the DD method. In both synthesis routes, control samples with no added amino acids were also prepared.

Samples for the NPD experiments were prepared as described above, but because of the higher amounts of material needed for these experiments, 4 times the amounts of solution (*i.e.*, 200 ml) were used in each case.

*Magnetization measurements*: Magnetic measurements were carried out at the Quantum Matter Research (QMR) center in the Department of Physics at the Technion. A few milligrams of each MCO sample were loaded into a plastic holder, which was placed inside a brass tube. The tube was placed inside the SQUID magnetometer (MPMS3, Quantum Design) and the system was sealed and purged with helium. Each sample underwent two types of measurements: magnetic moment *vs.* magnetic field (M−H) at a constant temperature (300 K and 2 K), and magnetic moment *vs.* temperature (M−T) at a constant magnetic field (10 Oe), while heating from 2 K up to 75 K and after cooling to 2 K at zero magnetic field.

*Amino acid analysis (AAA)*: As first step, AAs were removed from crystal surfaces by the addition of 1 ml of 1 M NaOH solution to several milligrams of a sample within an Eppendorf tube. The tube was sealed and agitated for 30 s, and the samples were then centrifuged and washed twice with DI water and twice with acetone. AAA was performed in Xell AG (Bielefeld, Germany). Each sample was dissolved in a weak HCl solution, and this was followed by a derivatization process with AQC (6-aminoquinolyl-N-hydroxysuccinimidyl carbamate).[65,66] The samples were injected into an ultra-high-performance liquid chromatograph (UHPLC, Agilent 1290) and AAs were detected using a UV (DAD detector).

*High-resolution powder X-ray diffraction*: HR-PXRD was carried out on beamline ID22 of the European Synchrotron Radiation Facility (ESRF) in Grenoble, France.[67] Several milligrams of each sample were transferred into borosilicate capillaries (0.5−0.7 mm in diameter), which were sealed using wax. Beam energy during the measurement was set at 35 keV. All samples were measured at room temperature, while selected samples were measured at 5 K, using liquid helium. Several samples were also measured after ex-situ annealing treatment (200ºC for 4 h, under vacuum). Rietveld refinement was performed on each diffractogram using the GSAS-II software,[68] and the lattice parameters of each sample, as well as diffraction peak width parameters, were extracted.

*Neutron powder diffraction*: Room temperature (300 K) and low temperature (3 K) NPD experiments were undertaken on the KANDI-II powder diffractometer at the Israeli Research Reactor II (IRR-II) at the Nuclear Research Centre – Negev (NRCN) [69] a monochromatic beam with a wavelength of $\lambda=2.47(1)$ Å. Powdered samples were loaded into vanadium cylindrical sample holders, each of 10-mm diameter and ~0.2 mm wall thickness. Diffraction patterns were analyzed using Rietveld analysis,[70] employing the FULLPROF code.[71] Low temperature was achieved using a closed cycle helium refrigerator.

*Scanning Electron Microscopy*: HR-SEM images were obtained with a Zeiss Ultra+ high-resolution scanning electron microscope (SEM). The primary electron energy was set to 1.5 keV for all samples; hence, no conductive coating was needed.

**Supporting Information**

Supporting Information is available online or from the author.

**Acknowledgements**

The authors thank Dr. Anna Eyal from the Quantum Matter Research (QMR) Center in the Physics Department, Technion, Haifa, for help with the magnetic measurements.

Diffraction experiments were performed on beamline ID22 (experiment MA 4534) at the European Synchrotron Radiation Facility (ESRF), Grenoble, France

A.L. acknowledges financial support by the Israeli Ministry of Energy, as part of the scholarships program for M.Sc. and Ph.D. students.

**Conflict of Interest**

The authors declare no conflict of interest.

**Supporting Information**

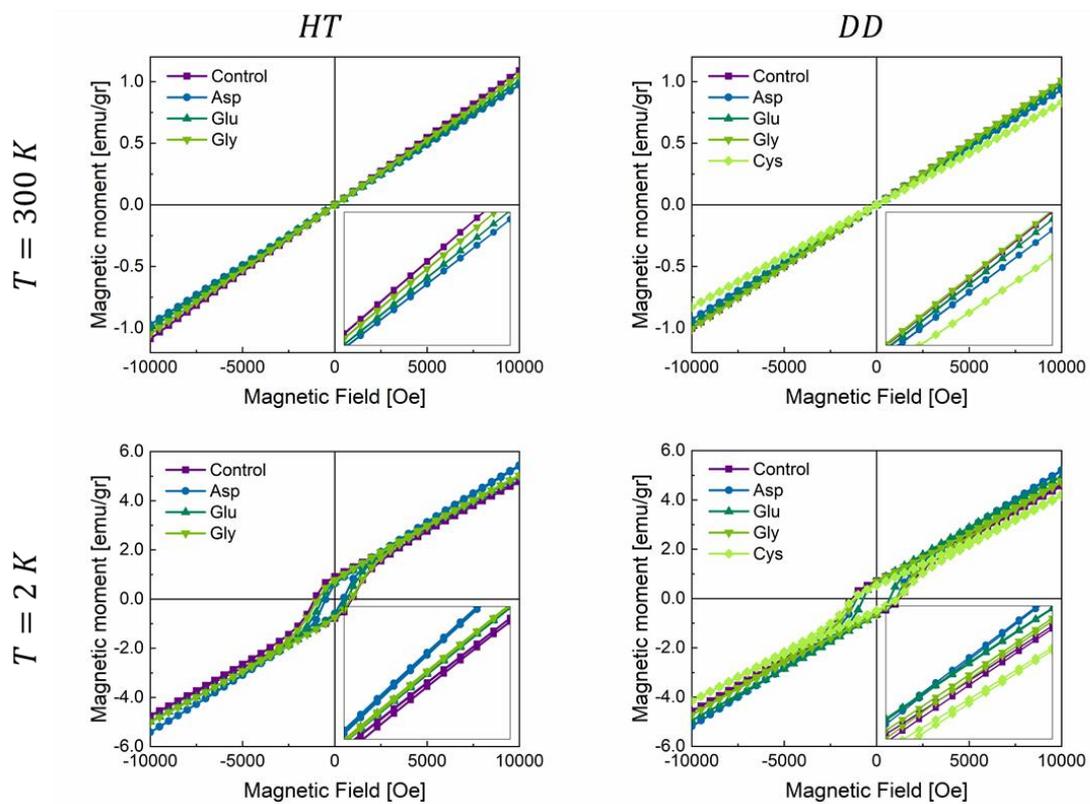

**Figure S1.** M−H plots of MCO grown with the same concentration (0.08 M in solution) of each of the four studied amino acids.

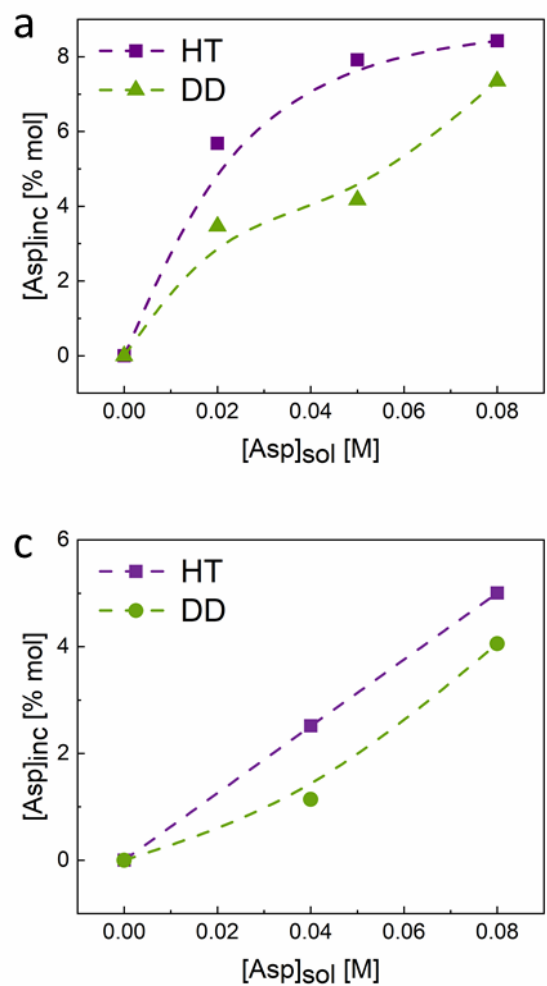

**Figure S2.** Concentration of incorporated Asp...in samples measured by PXRD (a) and by NPD (b) *vs.* their initial concentrations in the synthesis solution, as measured by AAA. Error bars are smaller than data points.

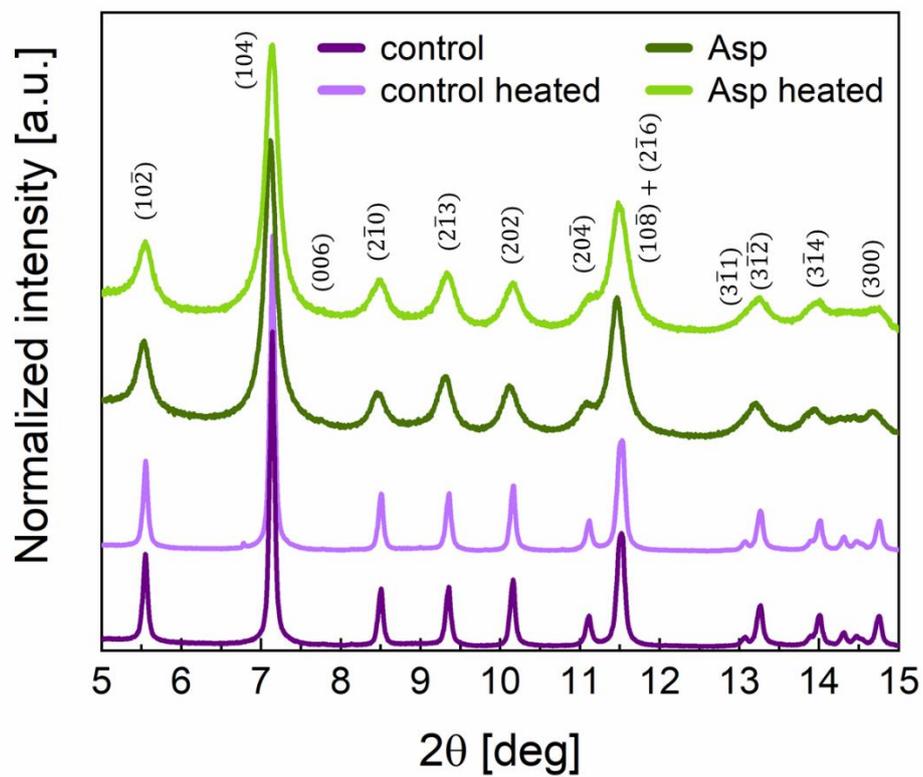

**Figure S3.** HR-PXRD measurements of pure and Asp-incorporated MCO, before and after thermal annealing.

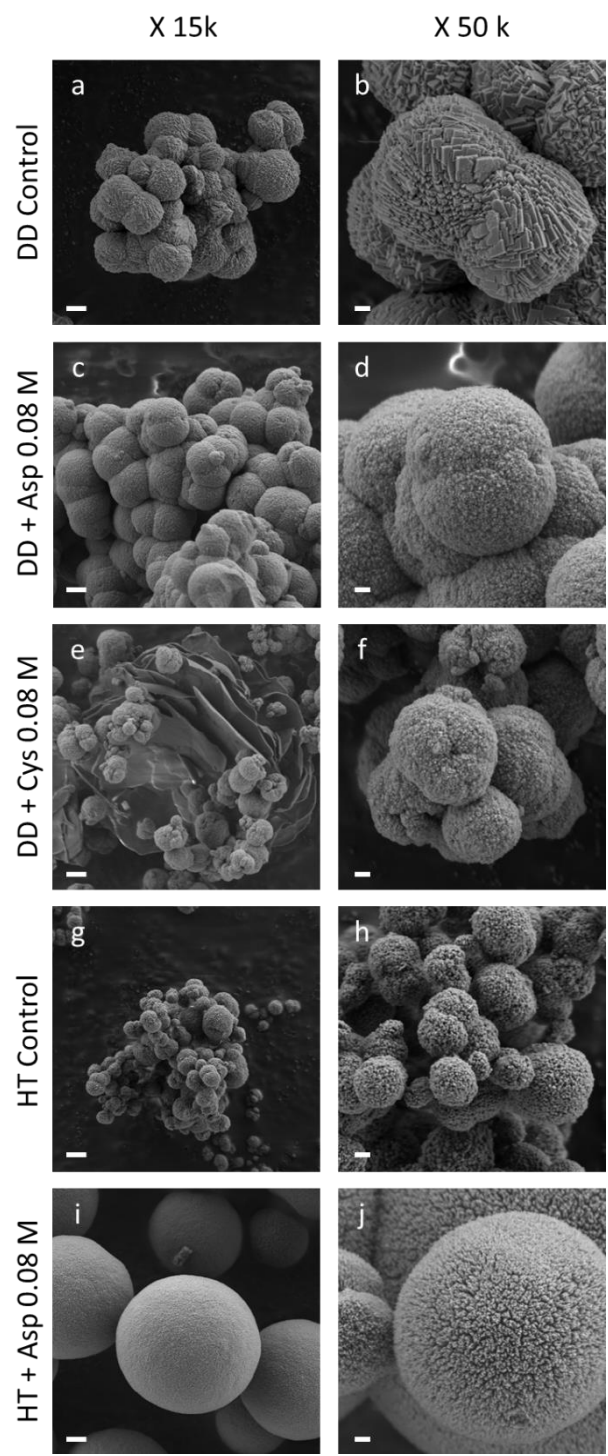

**Figure S4.** HR-SEM images of (a+b) DD-grown pure MCO; (c+d) DD-grown Asp-incorporated MCO; (e+f) DD-grown Cys-incorporated MCO; (g+h) HT-grown pure MCO; (i+j) HT-grown Asp-incorporated MCO. Left column: magnification of 15k (Scale bars: 1 µm); right column: magnification of 50k (Scale bars: 200 nm).

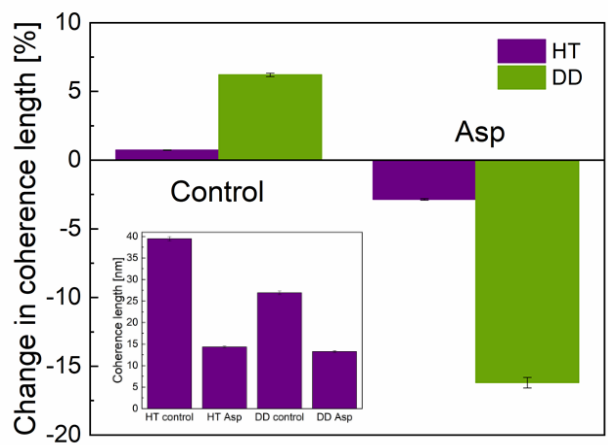

**Figure S5.** Change in coherence length of MCO due to thermal annealing. Inset: absolute coherence length.

**Table S1.** Rietveld refinement results for lattice parameters.

| Growth method | Heated | AA | Conc. [M] | $a^*$ [Å] | $da^*$ [Å] | $c^\#$ [Å] | $dc^\#$ [Å] | wR[†] [%] | GoF[‡] |
|---|---|---|---|---|---|---|---|---|---|
| **HT** | No  | Control | 0.00 | 4.782864 | 0.000150 | 15.67378 | 0.000383 | 6.52 | 3.22 |
| **HT** | No  | Asp     | 0.02 | 4.790165 | 0.000460 | 15.72787 | 0.001073 | 5.98 | 3.62 |
| **HT** | No  | Asp     | 0.05 | 4.793083 | 0.000541 | 15.74333 | 0.001241 | 5.74 | 3.14 |
| **HT** | No  | Asp     | 0.08 | 4.796713 | 0.000570 | 15.75922 | 0.001320 | 6.46 | 3.82 |
| **HT** | Yes | Control | 0.00 | 4.781319 | 0.000150 | 15.66930 | 0.000383 | 6.71 | 3.82 |
| **HT** | Yes | Asp     | 0.08 | 4.784520 | 0.000519 | 15.73200 | 0.001207 | 5.61 | 3.91 |
| **DD** | No  | Control | 0.00 | 4.793836 | 0.000263 | 15.70288 | 0.000644 | 6.47 | 4.23 |
| **DD** | No  | Asp     | 0.02 | 4.794053 | 0.000400 | 15.72491 | 0.000976 | 7.95 | 5.10 |
| **DD** | No  | Asp     | 0.05 | 4.795800 | 0.000426 | 15.72929 | 0.001024 | 7.55 | 3.61 |
| **DD** | No  | Asp     | 0.08 | 4.798733 | 0.000706 | 15.73384 | 0.001557 | 5.57 | 2.57 |
| **DD** | Yes | Control | 0.00 | 4.782802 | 0.000682 | 15.66308 | 0.000587 | 5.20 | 2.96 |
| **DD** | Yes | Asp     | 0.08 | 4.784779 | 0.000750 | 15.70120 | 0.001628 | 4.63 | 2.21 |
| **DD** | No  | Cys     | 0.02 | 4.791696 | 0.000637 | 15.70716 | 0.001496 | 9.54 | 4.41 |
| **DD** | No  | Cys     | 0.05 | 4.790649 | 0.000719 | 15.70751 | 0.001688 | 10.6 | 5.90 |
| **DD** | No  | Cys     | 0.08 | 4.791742 | 0.001398 | 15.70817 | 0.003098 | 11.5 | 7.07 |

[*] $a$, $da$ – value and error in $a$ lattice parameter.  
[#] $c$, $dc$ – value and error in $c$ lattice parameter.  
[†] wR - weighted profile R-factor.  
[‡] GoF - goodness-of-fit parameter

**Table S2.** Rietveld refinement results for line broadening.

| Growth method | Heated | AA | Conc. [M] | $L^*$ [nm] | $dL^*$ [nm] | $S^\#$ [] | $dS^\#$ [] | $wR^\dagger$ [%] | $GoF^\ddagger$ |
|---|---|---|---|---|---|---|---|---|---|
| **HT** | No | Control | 0.00 | 39.47 | 0.5 | 3416.4 | 127.5 | 6.52 | 3.22 |
| **HT** | No | Asp | 0.08 | 14.36 | 0.2 | 10350.4 | 489.4 | 6.46 | 3.82 |
| **HT** | Yes | Control | 0.00 | 39.76 | 0.5 | 3409.7 | 126.7 | 6.71 | 3.82 |
| **HT** | Yes | Asp | 0.08 | 13.95 | 0.2 | 10765.3 | 447.4 | 5.61 | 3.91 |
| **DD** | No | Control | 0.00 | 26.92 | 0.4 | 6758.2 | 229.3 | 6.47 | 4.23 |
| **DD** | No | Asp | 0.08 | 13.28 | 0.2 | 13369.4 | 588.1 | 5.57 | 2.57 |
| **DD** | Yes | Control | 0.00 | 28.59 | 0.4 | 9344.3 | 210.4 | 5.20 | 2.96 |
| **DD** | Yes | Asp | 0.08 | 11.13 | 0.2 | 12841.5 | 616.2 | 4.63 | 2.21 |

$^*$ *a*, d*a* – value and error in *a* lattice parameter.  
$^\#$ *c*, d*c* – value and error in *c* lattice parameter.  
$^\dagger$ wR - weighted profile R-factor.  
$^\ddagger$ GoF - goodness-of-fit parameter

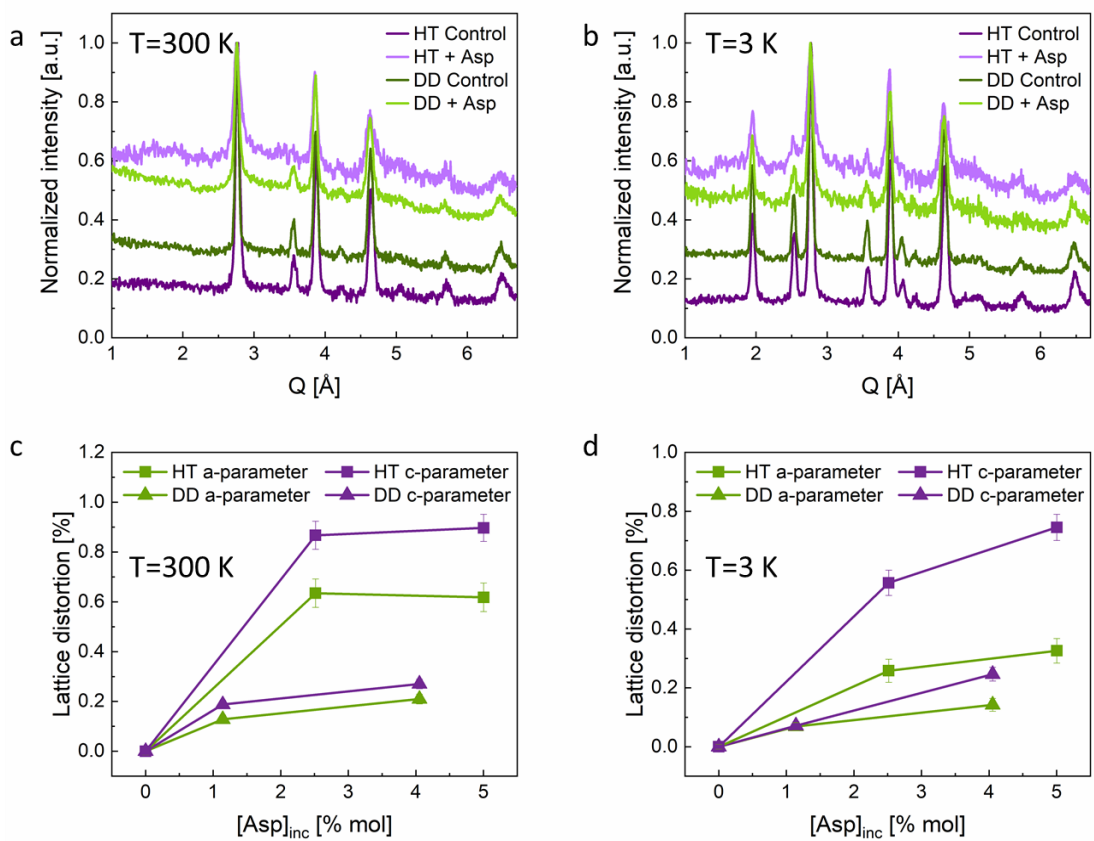

**Figure S6.** NPD assessment of pure and Asp-incorporated MCO (a) at room temperature and (b) at 3 K.. (c+d) Lattice distortion of MCO calculated using eq. (2), based on NPD data.

Neutron powder diffraction (NPD) was measured on MCO powder samples obtained both by DD and by HT synthesis routes. Diffraction data are consistent with rhombohedral symmetry. All RT patterns are well described by a model consisting of the *R-3c* space group with Mn in the *6b*(0,0,0), C in the *6a*(0,0,1/4), and O in the *18e*(x~0.28,0,1/4) positions. Lattice parameters values are in the vicinity of a~4.8, and c~15.7, in agreement with the XRD data. Control samples (*i.e.* with 0% Asp incorporation) show clear diffraction patterns with line width close to the instrumental resolution, and signal-to-background ratio of ~4 for the strongest reflection [(012) in the *R-3c* model; **Figure S6a**]. Samples containing Asp show significant broadening of reflections in agreement with the XRD results, and a large reduction in the signal-to-background ratio. For samples with an Asp molar concentration of 8% (in solution) this ratio reduces from ~5 to ~2 for the strongest reflection (bottom of **Figure S6a**).

NPD data of Asp samples at low temperature (LT, 3K) show reflections additional to those observed at RT. The positions of these additional reflections are consistent with the magnetic propagation vector k=(0,0,0) (*i.e.,* magnetic unit cell periodicity equals that of the crystallographic unit cell). Symmetry analysis of the possible magnetic structure was undertaken using the k-Subgroupsmag code in the Bilbao Crystallographic Server package.[72] The resulting structures were then used in Rietveld analysis of the data for all samples. Of the possible structures, the monoclinic *C2'/c'* (#15.89) in the basis (a,b,c;0,0,0) of the parent space group ***R-3c***, showed the best fit to the data for all measurements on all samples. This structure consists of the same lattice parameter constraints as in *R-3c*, and Mn in the (0,0,0), C in the (0,0,1/4), O1 in the (x~0.25, 0, ¼), and O2 in the (0, y~0.25, 1/4) positions.

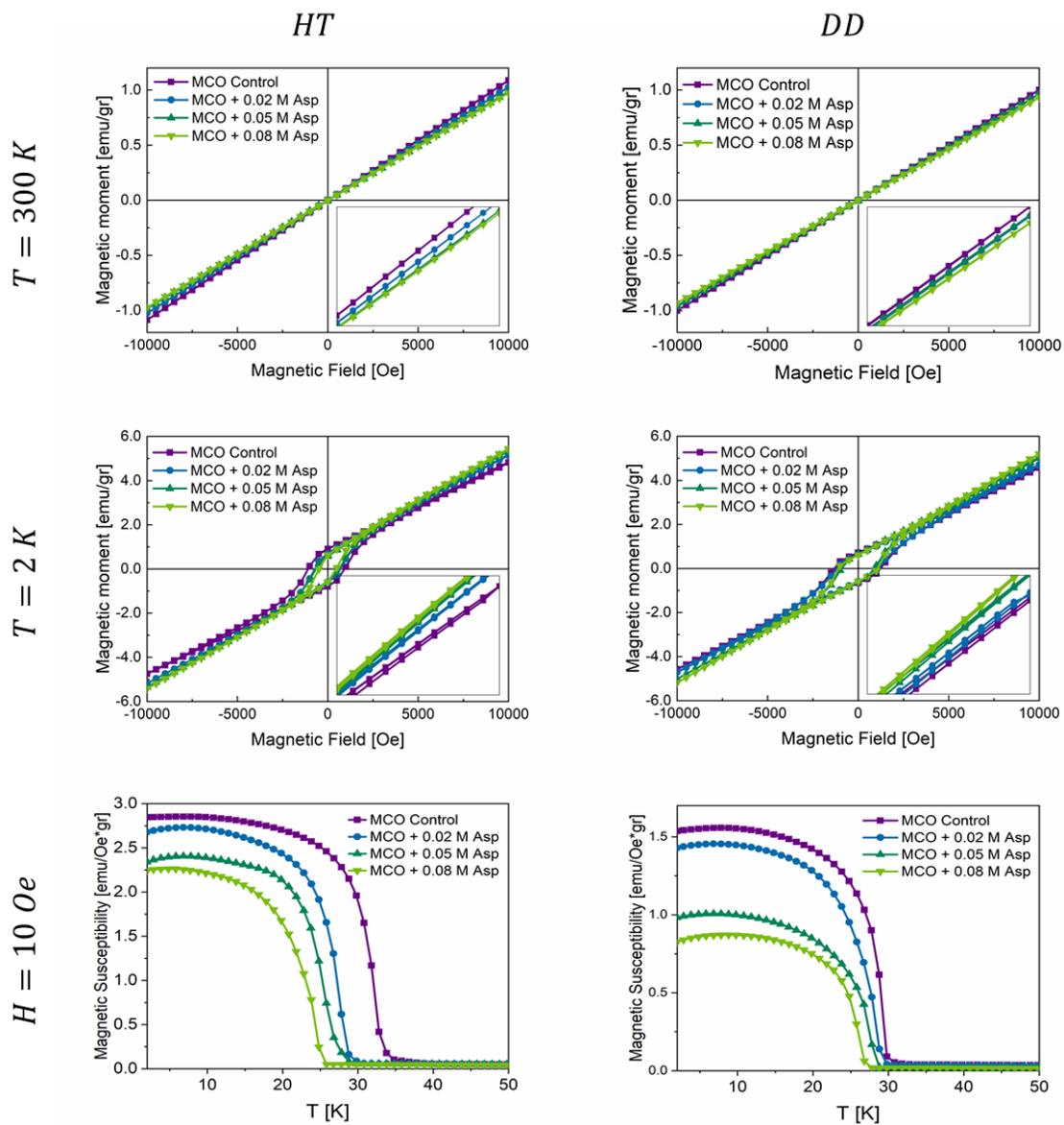

**Figure S7.** M−H and M−T curves of the different Asp-incorporated MCO samples.

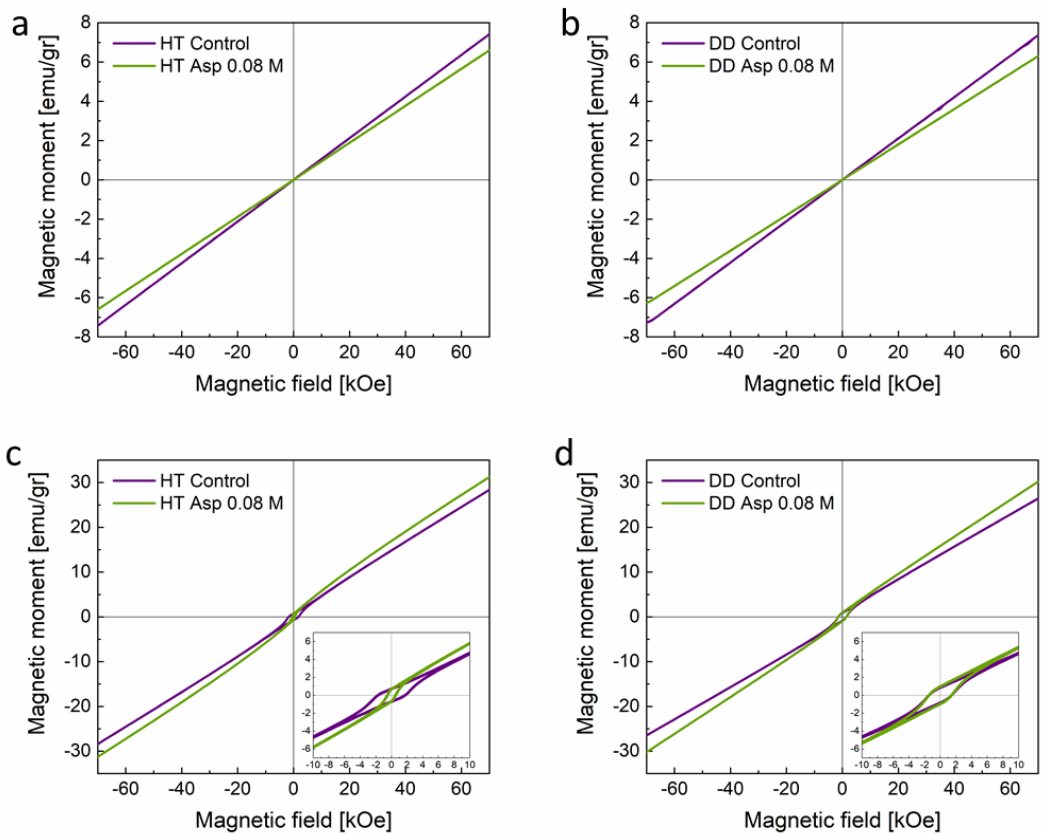

**Figure S8.** M−H curves of control and Asp-incorporated MCO at high fields (−7 to 7 T). (a+b), measured at 300 K; c+d), measured at 2 K.

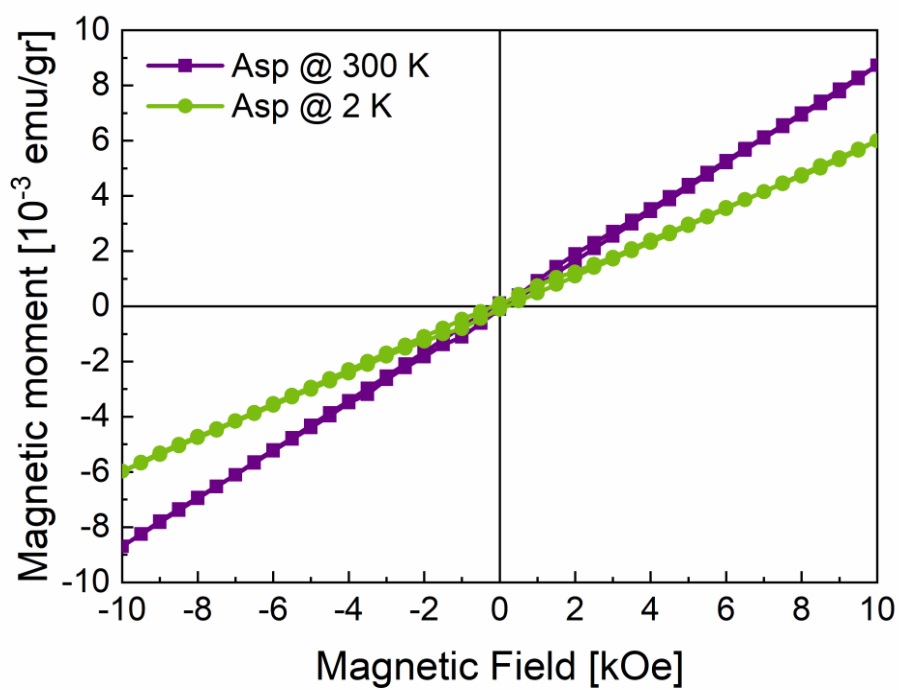

**Figure S10.** M−H curves of Asp powder.

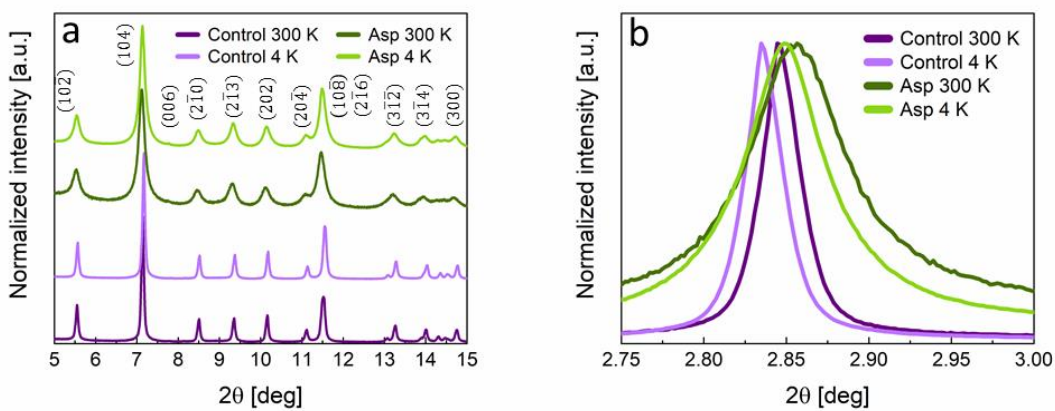

**Figure S11.** a) HR-PXRD of control and Asp-incorporated MCO, measured both at room temperature and at T=5 K. b) Shifts of the (104) reflections are due to cooling.

**Table S3.** Rietveld refinement results for cryo-HR-PXRD samples, measured at T = 5 K.

| Growth method | [Asp] [M] | $a^*$ [Å] | $da^*$ [Å] | $c^\#$ [Å] | $dc^\#$ [Å] | $x^\S$ | $dx^\S$ | $wR^\dagger$ [%] | $GoF^\ddagger$ |
|---|---|---|---|---|---|---|---|---|---|
| HT | 0.00 | 4.776199 | 0.000160 | 15.594803 | 0.000412 | 0.27168 | 0.00032 | 7.96 | 7.41 |
| HT | 0.02 | 4.787685 | 0.000518 | 15.656055 | 0.001220 | 0.27054 | 0.00034 | 6.77 | 2.35 |
| HT | 0.05 | 4.789975 | 0.000350 | 15.677828 | 0.000809 | 0.26930 | 0.00028 | 6.48 | 2.71 |
| HT | 0.08 | 4.792005 | 0.000409 | 15.685464 | 0.000940 | 0.26924 | 0.00046 | 7.40 | 7.37 |
| HT | 0.00 | 4.780700 | 0.000397 | 15.605459 | 0.000845 | 0.27030 | 0.00109 | 2.93 | 1.38 |
| HT | 0.02 | 4.780674 | 0.000468 | 15.631359 | 0.001005 | 0.26988 | 0.00058 | 4.23 | 2.16 |
| DD | 0.05 | 4.789476 | 0.000258 | 15.657504 | 0.000625 | 0.26926 | 0.00041 | 10.28 | 4.50 |
| DD | 0.08 | 4.792866 | 0.000264 | 15.670914 | 0.000641 | 0.26905 | 0.00065 | 3.85 | 1.82 |

\* $a$, $da$ – value and error in $a$ lattice parameter.  † wR - weighted profile R-factor.

\# $c$, $dc$ – value and error in $c$ lattice parameter.  ‡ GoF - goodness-of-fit parameter

\# $x$, $dx$ – value and error in the relative oxygen position.

**The case of Cysteine**

After studying **the effects** of Asp incorporation on the magnetic properties of MCO, we wanted to study the same in the case of Cys-incorporation, which initially showed a high effect on susceptibility (Figure 1). AAA revealed high levels of Cys incorporation with a maximum value of ~8 mol %. (**Figure S2b**). On the other hand, HR-PXRD did not reveal any significant changes in the lattice parameter of MCO upon Cys incorporation (**Table S1**). The formation of additional crystalline phases was clearly observed (**Figure S11a**). The most prevalent among them, other than the major MCO phase, is α-$Mn_2O_3$ (bixbyite, see **Figure S11a**),[73] which increases in amount as the Cys concentration in the solution is increased (**Figure S11b**). This new phase can be also observed using HR-SEM as micron-sized plates (**Figure S4e**). One possible explanation for the Mn$_2$O$_3$ formation is the oxidation of Mn$^{+2}$ to Mn$^{+3}$ by the rapidly formed Cystine in the solution.[74] When we measured the magnetism of the Cys-incorporated samples, the magnetic susceptibility was indeed found to decrease both at room temperature and at 2 K (**Figure S11d**). It seems that the reason for the change in this case was the formation of Mn$_2$O$_3$, which reduces the effective amount of the magnetic MCO, hence the overall magnetization was decreased. Note that while α-Mn$_2$O$_3$ is also antiferromagnetic below 80 K, no signs of transition were detected (**Figure S12**). This is probably due to its relatively low amount.

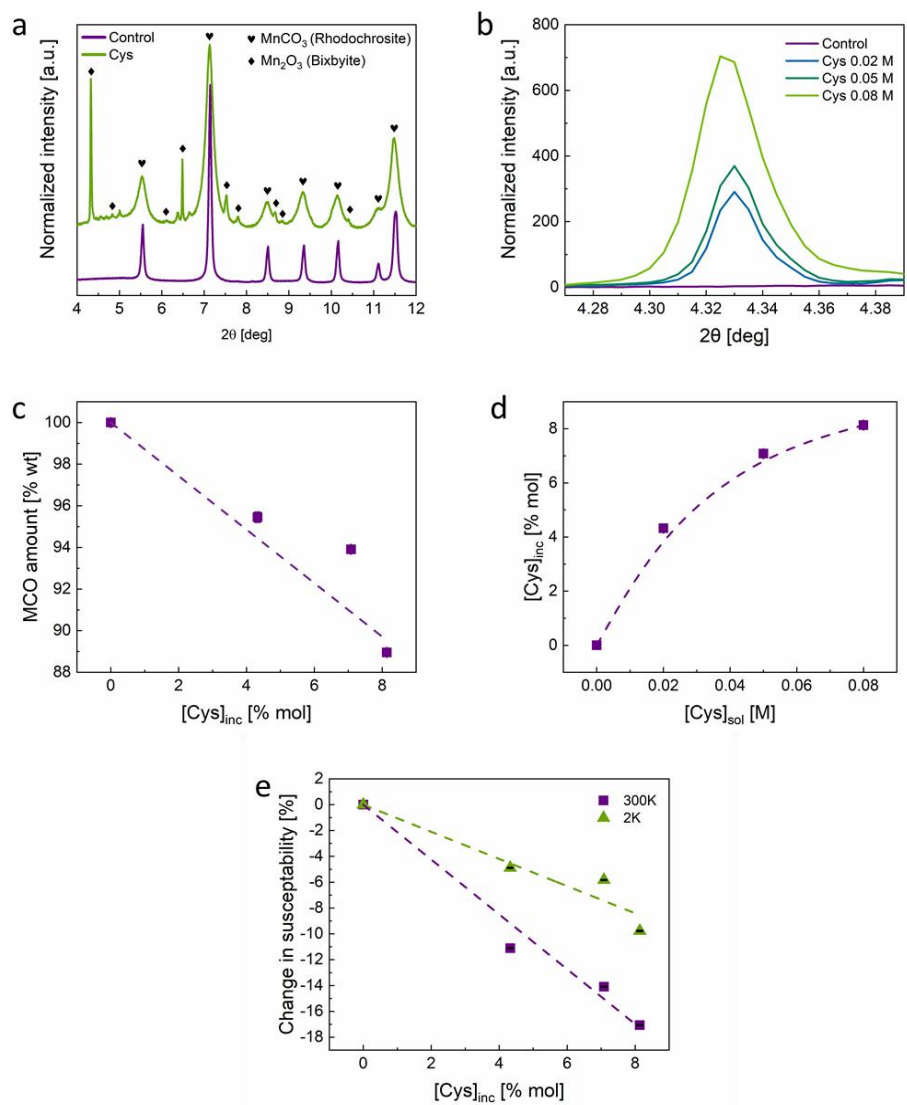

**Figure S11. Analysis of Cys-incorporated MCO.** (a) HR-PXRD of MCO grown with and without Cys, showing the formation of $Mn_2O_3$. (b) Evolution of the $Mn_2O_3$ (200) reflection with increasing amounts of Cys in solution. (c) decrease in relative amount of MCO due to the formation of $Mn_2O_3$, as calculated using Rietveld refinement. (d) AAA of Cys-MCO samples. (e) Decrease in magnetic susceptibility as a function of incorporated Cys, measured both at 300 K and at 2 K. Error bars are smaller than the data points.

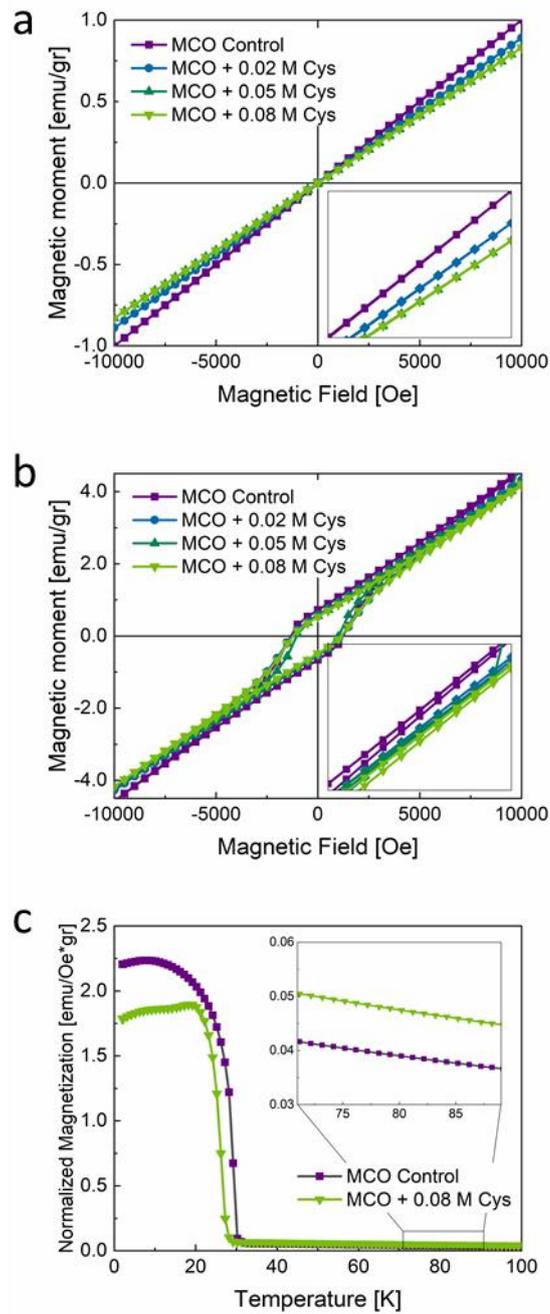

**Figure S12.** M−H and M−T curves of the different Cys-incorporated MCO samples. The inset in (c) shows that no AFM transition is observed for $Mn_2O_3$.